\documentclass[10pt,conference]{IEEEtran}

\usepackage[square, numbers]{natbib}
\usepackage{graphicx}
\usepackage{amsmath}
\usepackage{amssymb}
\usepackage{color}
\usepackage{url}
\usepackage{multirow}
\usepackage[linesnumbered,ruled]{algorithm2e}
\usepackage{verbatim}
\usepackage{subfigure}

\def\>{\ensuremath{\rangle}}
\def\<{\ensuremath{\langle}}

\def\0{\ensuremath{|0\rangle}}
\def\1{\ensuremath{|1\rangle}}

\usepackage{etoolbox}
\makeatletter
\patchcmd{\@makecaption}
  {\scshape}
  {}
  {}
  {}
\makeatother

\begin{document}

\title{Quantum Supremacy Circuit Simulation on Sunway TaihuLight}

\author{Riling Li, Bujiao Wu, Mingsheng Ying, Guangwen Yang, Xiaomin Sun}

\author{\IEEEauthorblockN{1\textsuperscript{st} Riling Li}
\IEEEauthorblockA{\textit{Department of Computer Science \& Technology}\\
\textit{Tsinghua University}\\
Beijing, China \\
rl-li16@mails.tsinghua.edu.cn}
\and
\IEEEauthorblockN{2\textsuperscript{nd} Bujiao Wu}
\IEEEauthorblockA{\textit{CAS Key Lab of Network Data Science and Technology} \\
\textit{Institute of Computing Technology, Chinese Academy of Sciences} \\
Beijing, China \\
wubujiao@ict.ac.cn}
\and
\IEEEauthorblockN{3\textsuperscript{rd} Mingsheng Ying}
\IEEEauthorblockA{\textit{Centre for Quantum Software and Information, University of Technology Sydney}\\
Sydney, Australia \\ \textit{State Key Lab of Computer Science, Institute of Software, Chinese Academy of Sciences}\\ \textit{Department of Computer Science \& Technology, 
Tsinghua University}\\ Beijing, China\\ 
Mingsheng.Ying@uts.edu.au}
\and
\IEEEauthorblockN{4\textsuperscript{th} Xiaoming Sun}
\IEEEauthorblockA{\textit{CAS Key Lab of Network Data Science and Technology} \\
\textit{Institute of Computing Technology, Chinese Academy of Sciences} \\
Beijing, China \\
sunxiaoming@ict.ac.cn}
\and
\IEEEauthorblockN{5\textsuperscript{th} Guangwen Yang}
\IEEEauthorblockA{\textit{Department of Computer Science \& Technology}\\
\textit{Tsinghua University}\\
Beijing, China \\
ygw@tsinghua.edu.cn}
}

\maketitle

\begin{abstract}
\quad With the rapid progress made by industry and academia, quantum computers with dozens of qubits or even larger size are being realized. However, the fidelity of existing quantum computers often sharply decreases as the circuit depth increases. Thus, an ideal quantum circuit simulator on classical computers, especially on high-performance computers, is needed for benchmarking and validation. We design a large-scale simulator of universal random quantum circuits, often called \textquotedblleft quantum supremacy circuits\textquotedblright, and implement it on Sunway TaihuLight. The simulator can be used to accomplish the following two tasks: 1) Computing a complete output state-vector; 2) Calculating one or a few amplitudes. We target the simulation of 49-qubit circuits. For task 1), we successfully simulate such a circuit of depth 39, and for task 2) we reach the 55-depth level. To the best of our knowledge, both of the simulation results reach the largest depth for 49-qubit quantum supremacy circuits.
\end{abstract}


\section{Introduction}\label{section1}

\quad The concept of quantum computer was proposed almost four decades ago \cite{feynman1982}. But until recently it had been unknown whether quantum computers can indeed exceed the computing capability of their classical predecessors. Thanks to the progress made by industry and academia in recent years, practical quantum computing might become reality soon. However, before a commercial quantum computer is launched on market, many tests and verification need to be done. One of the most important is to test the fidelity of quantum circuit. One way to accomplish this is to simulate quantum circuits by computing the ideal state amplitudes on a classical computer.  A quantum simulator on classical computer could also be used to verify correctness of certain  quantum algorithms and benchmark the notion ``quantum supremacy".

\par
\begin{figure}
    \centering
    \includegraphics[scale=0.5]{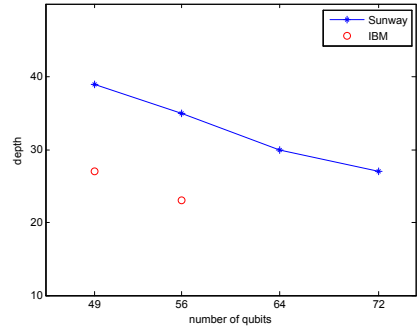}
    \caption{The maximal circuit depth we can simulation in cases of different number of qubits. Both our simulator and IBM \cite{ibm} can calculate all $2^{49}$ amplitudes for 49-qubit case. For 56- or more qubit cases, we only calculate a slice of $2^{32}$ amplitudes for purpose of demonstration.}
    \label{qubit_scaling}
\end{figure}

\setlength\parindent{1em} Quite a few implementations of quantum circuit simulators have been developed in last few years \cite{qhipster,liqui,svor,eth,ibm,googleSim}. To test the limitation of quantum circuit simulation on classical computers, we design a simulator on Sunway TaihuLight, one of the most powerful high-performance computer at present in the world. We mainly aim at 49-qubit circuits, because such circuits are hard to directly simulate on other supercomputers due to the limited memory spaces, and it is widely believed that near-term quantum device will first achieve quantum supremacy at this scale. Our simulator can accomplish the following: 
\begin{itemize} \item \textbf{\textit{Task 1}}: Computing a complete output state-vector representing a quantum state output by the simulated quantum circuit; 
\item \textbf{\textit{Task 2}}:  Sampling (i.e. calculating a small amount of) the amplitudes of a quantum state output by the quantum circuit. 
\end{itemize}
For Task 1, we are able to solve the lattice of $7\times7$ qubits with depth 39 \footnote{The first layer of Hadamard gates are not counted as the circuit depth, we treat it as layer 0. So the circuit we simulate is from layer 0 to layer 39. This also holds for Task 2.} in 4.2 hours using 131072 core groups, which is around 80\% of the computing resource of Sunway. As to Task 2, our method can calculate one amplitude for 49-qubit circuits of depth 55.
Moreover, our method for Task 1 can also be directly extended to the lattices of $7\times8$, $8\times8$, and $9\times 8$ qubits for calculating a few amplitudes, though the reachable depth of the random circuit would be decreasing with the increasing of the qubits. Figure \ref{qubit_scaling} shows the maximal depth our simulator can reach for different number of qubits.

\subsection{Comparison with other quantum simulators}

For Task 1, the simulator described in \cite{eth} can simulate a quantum circuit with $5\times9$ qubits and depth 25 on the Cori \uppercase\expandafter{\romannumeral2} supercomputer in less than 10 minutes, using 0.5 petabytes and 8192 nodes. Reference \cite{ibm} reported the simulation of a $7\times7$ grid of qubits with depth 27, which costs more than one day on IBM Blue Gene/Q. In contrast, our method simulates a $7\times7$ grid of qubits with depth 39 in much less time. The task of simulating $7\times8$ grid of qubits with depth 23 was not finished in \cite{ibm} because $2^{56}$ amplitudes are too many to calculate. For this task, they only calculated $2^{39}$ amplitudes. In contrast, our simulator can calculate $2^{37}\sim 2^{42}$ amplitudes in a short time for $7\times8$-qubit circuits of depth 35. Table \ref{comparison} compares our implementation and several previous works.

Task 2 of sampling amplitudes can be used to estimate the fidelity by computing the cross entropy, which usually needs $10^{3}\sim 10^6$ samples \cite{googlequantumsupremacy}. The sampling target is to calculate an amplitude $\alpha_{x}$, where $0\leq x\leq2^{n}-1$ for an $n$-qubit circuit. The method of calculating the state-vector can also be directly applied to calculate a small number of amplitudes, though in this way the reachable depth is not very large. However, we can get one amplitude for $7\times 7$-qubit circuits of depth 55 by calculating the inner product of two 49-qubit state-vector, although this is very expensive. Our results of Task 1 and Task 2 shows that the bound of 49-qubit with depth 40 in \cite{googleSim,googlequantumsupremacy} is no more out of reach.

\begin{table}[]
    \centering
    \begin{tabular}{|c|c|c|c|c|}
    \hline
     \textbf{Reference} & Platform & \textbf{ Qubits} & \textbf{ Depth} & \textbf{ Time} \\
    \hline
    \hline
       ETH\cite{eth} & Cori \uppercase\expandafter{\romannumeral2} & 45 & 25 & 10 minutes\\
    \hline
       IBM\cite{ibm} & Blue Gene/Q & 49 & 27 & 2 days \\
    \hline
       Sunway & Sunway TaihuLight & 49 & 39 & 4.2 hours\\
    \hline
    \end{tabular}
    \caption{Several quantum circuit simulators implemented on supercomputers in recent years. Our simulator reaches the largest number of qubits and also the largest depth for 49-qubit circuits in the case of solving all amplitudes of the final state (Task 1).}
    \label{comparison}
\end{table}

\subsection{Technical contributions of this paper}

The first contribution of this paper is that we introduce a novel partition scheme via a dynamic programming algorithm, by which we can save more time and space than \cite{ibm}. At the same time, we propose several \textit{global} optimizing techniques, including some new optimizations adaptive to the structure of 2D grid circuits and several other optimizations for reducing the network communication amount when implementing our method.

Our second contribution is some new \textit{local} optimizing techniques on a single node, which take the advantage of the many-core heterogeneous processor of Sunway. Our optimization greatly reduces the amount of memory access while it only increases a small quantity of calculation. We also apply some standard optimization such as vectorization. By all of these techniques, we can simulate a 28-qubit circuit on one core group very quickly, which is significant for improving the performance of 49-qubit circuit simulation.

Sunway TaihuLight adopts a popular many-core heterogeneous system architecture, of which in one CPU chip there are 4 core groups (CGs) each with 1 management processing elements (MPEs) and 64 computing processing elements (CPEs) \cite{sunway2016}. Since in our implementation each core group corresponds to an unique MPI process, we regard a core group instead of a CPU chip as a single node in this paper to avoid some messy and confusing description. That is, \emph{each node contains only 1 master core (MPE) and 64 slave cores (CPEs) and corresponds to only one MPI process.}

\subsection{Organisation of the paper} Section 2 describe the simulation methodologies and optimization techniques. Section 3 presents the numerical results of our implementation as well as verification of the results. Section 4 draws a conclusion and discusses the problems to be solved in the future work.

\section{Methodologies and optimizations}

Before describing our methods and optimizations, let us recall some basic notions and notations. 

A basic storage unit in a quantum computer is a quantum bit (qubit). Generally, we can use a $2^{n}$-length complex vector to describe an $n$-qubit state, such as $|\psi\> = (\alpha_0,\alpha_1,...\alpha_{2^n-1})^T$. A practical quantum circuit consists of single-qubit and 2-qubit gates. For example, a single-qubit gate $U^{k}=\begin{pmatrix}
a & b\\
c & d
\end{pmatrix}
$ on qubit $k$ can be treated as an $n$-qubit gate which acts as identity on the other $n-1$ qubits, as denoted by $U = I^{\otimes n-k-1}\otimes U^{k} \otimes I^{\otimes k}$, where $I$ is the identity operator on a single qubit. In this paper, a superscript indicates the qubit that the gate is performed on, and a subscript is used as a gate label or an index of amplitude.

To perform $U^{k}$ on a state $|\psi\> = (\alpha_0,\alpha_1,\cdots,\alpha_{2^n-1})^T$, we have:
$$
\begin{pmatrix}
\alpha'_i\\ 
\alpha'_{i+2^k}
\end{pmatrix}
=
\begin{pmatrix}
a & b\\ 
c & d
\end{pmatrix}
\begin{pmatrix}
\alpha_i\\ 
\alpha_{i+2^k}
\end{pmatrix}
$$
for every $i = i_{n-1}...i_{1}i_{0}$ where $i_k = 0$. And the resulting state is then $|\psi'\> = (\alpha'_0,\alpha'_1,...\alpha'_{2^n-1})^T = U^k|\psi\>$. The 2-qubit gates used in this paper are mainly the controlled-gate: CZ. A controlled-gate $CU^{c,t}$ act on qubits $c$ and $t$, with the first being the control bit and the second being the target bit. The performance of a 2-qubit gate is similar to that of a single-qubit gate with the only extra consideration of whether the control qubit is in state $|1\>$; for more details, we refer to \cite{qiqc}.

\par
\begin{figure*}
    \centering
    \includegraphics[scale=0.4]{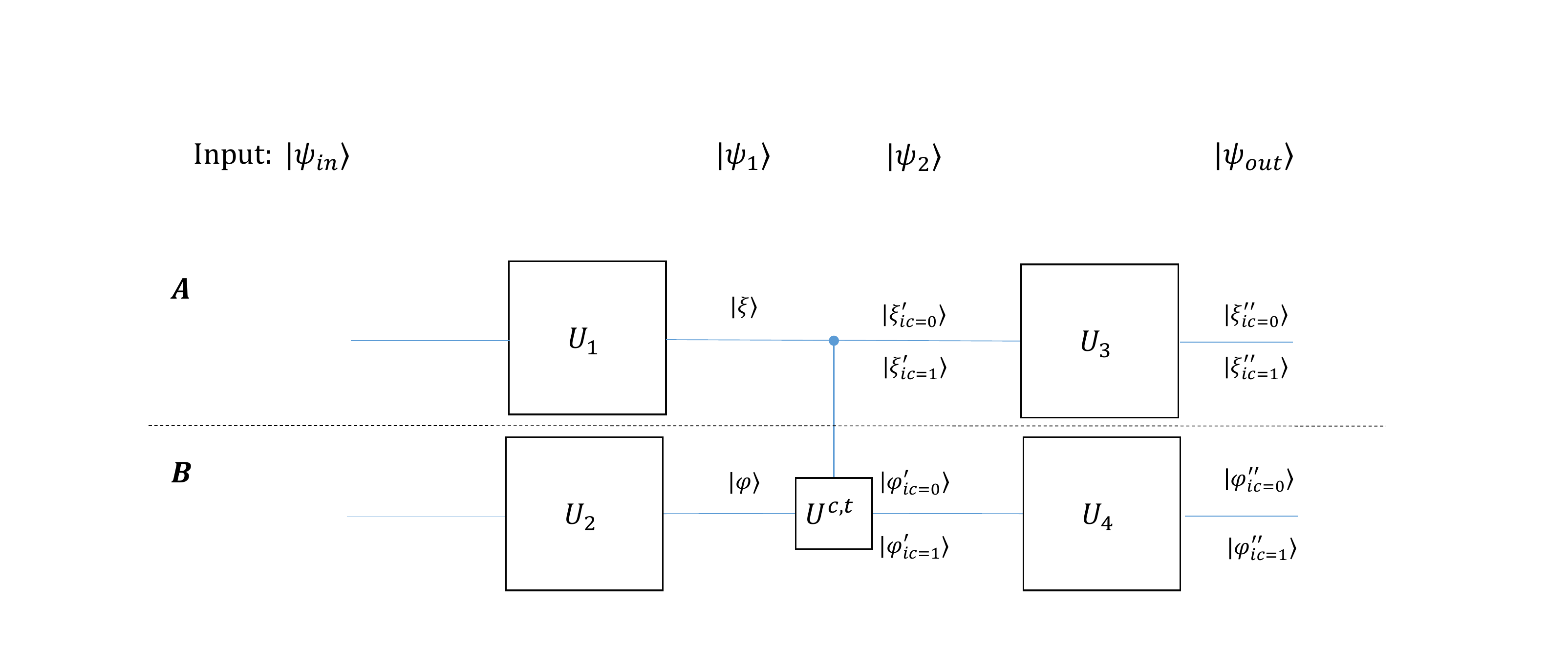}
    \caption{An illustrative example of how gate decomposition works. The initial state is $|\psi_{in}\>=|0\>^{\otimes n}$ and the circuit is $\mathcal{U}_{circuit} = (U_3 \otimes U_4)CU^{c,t}(U_1 \otimes U_2)$. The whole system can be regarded as a composition of two subsytems \textbf{A} and \textbf{B}, while the gate $CU^{c,t}$ (qubit $c$ in \textbf{A} and qubit $t$ in \textbf{B}) causes their entanglement. Assume \textbf{A} has $n_A$ qubits and \textbf{B} has $n_B$ qubits, and the total qubit number is $n=n_A+n_B$. After performing $U_1$ and $U_2$, the whole state is $|\psi_1\>=|\xi\>\otimes|\varphi\>$. With the decompsing of $CU^{c,t}$ we get two branches: qubit $c$ at $\0$ and qubit $c$ at $\1$. After performing $CU^{c,t}$, we have $|\xi^{'}_{i_c}\>$ and $|\varphi^{'}_{i_c}\>$, $i_c = 0,1$. Finally, we perform $U_3$, $U_4$ and get $|\xi^{''}_{i_c}\> = U_3|\xi^{'}_{i_c}\>$, $|\varphi^{''}_{i_c}\> = U_4 |\varphi^{'}_{i_c}\>$. The result is $|\psi_{out}\> = \sum_{i_c=0,1}|\xi^{''}_{i_c}\>\otimes|\xi^{''}_{i_c}\>$ . Note that after performing $CU^{c,t}$ we need to double the space to storage the two branches (qubit $c$ at $\0$ or $\1$). So, the total space consumption is $2^{n_A+1}+2^{n_B+1}$, in contrast to $2^n$, the space consumption of directly calculating the resulting state of the composite system gate by gate.}
    \label{circuit_ex1}
\end{figure*}

\setlength\parindent{1em} For a quantum circuit simulation, the initial state is usually the product state: 
$$|0\>^{\otimes n} = \underset{n}{\underbrace{\0\otimes \0 \otimes...\otimes \0}}.$$
If there is no 2-qubit gate in the circuit, the state will persistently remain a product state and only $O(n)$ space is needed to describe the state. But as the number of two-qubit gates increases, the quantum state may become highly entangled. In this case, the storage of the state will require $O(2^{n})$ space. As $n$ increases, the memory required becomes too large even for the most powerful supercomputers. For example, the maximal qubit number of a state vector that could be stored in the memory of Sunway is 45 (46) using double (float), which requires 0.5 petabytes of memory space.
However, for a 2-qubit gate $CU^{c,t}$, we can decompose it to $CU^{c,t} = P_{0}^{c}\otimes I^{t} + P_{1}^{c}\otimes U^{t}$, where $P_0 = |0\>\<0|$ and $P_1=|1\>\<1|$, and go along two branching path with respect to $P_0$ and $P_1$. deferring the entanglement it brings. Until an appropriate stage we combine the branching paths, finishing the deferred entanglement. Essentially, this idea comes from the notion of ``Feynman path integral" and appeared in \cite{chenlijie,ibm,googleSim}. Figure \ref{circuit_ex1} gives a simple example showing how circuit partition works.

A universal random quantum circuit has a 2D grid architecture \cite{googlequantumsupremacy}. The quantum gates used in this type of circuits are $H,X^{1/2},Y^{1/2},T$ and CZ. The 2-qubit gate CZ only appears between two adjacent qubits in the grid. Since in each layer of an $n\times m$-grid universal random circuit the positions of CZ gates are fixed, we can find a concrete partition scheme according to the scale of circuit to be simulated. Our techniques for finding this partition scheme are described in detail in the following subsection.

\subsection{Method for computing the complete state-vector (Task 1)}\label{sec21}
 \quad Figure \ref{circuit_ex1} describes the basic ideas of circuit partition. But as the circuit depth increases, the number of decomposed 2-qubit gates also increases. To increase the depth of circuits that can be simulated, we propose two optimizing techniques, which enable us to simulate 49-qubit circuits of depth 39, computing the complete output state-vector: 
\begin{itemize}
    \item \textbf{\textit{Technique 1}}: We analyze the structure of universal random circuits, exploit the diagonal properties of CZ gates, and propose a technique, called \textit{implicit decomposition}, which can decompose extra 7 CZ gates without requiring too much extra memory space, so as to increasing the depth of circuits that could be simulated by 8.
    \item \textbf{\textit{Technique 2}}: We propose a dynamic programming (DP) algorithm to find a good partition scheme for a given simulation task. In contrast to the general heuristic search method that usually takes a long time, this DP algorithm is efficient and can find an optimal partition scheme under certain constrains. It also makes the simulation easier to optimize and parallelize and thus improves the performance in time-to-solution.
\end{itemize}

\subsubsection{Implicit decomposition}
Our partition scheme divides the target circuit into three parts, and each part requires less memory than the memory needed to store the entire 49-qubit state vector. Implicit decomposition balances the memory requirement of these 3 parts, decomposes extra CZ gates and increases the depth of circuits that could be simulated. 

\begin{figure*}
    \centering
    \includegraphics[scale=0.3]{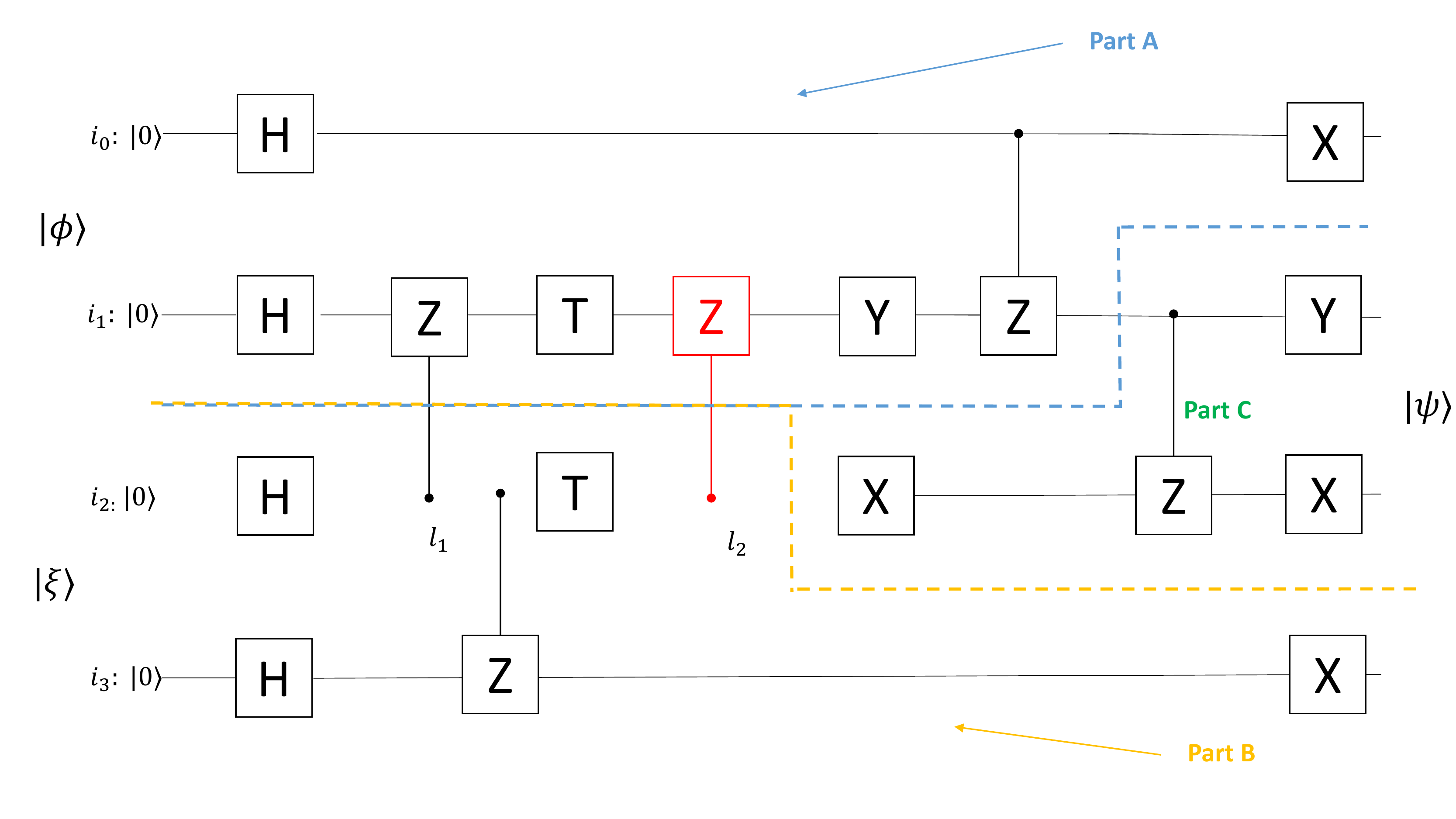}
    \caption{Example of a 4-qubit circuit. In this example, there are 2 CZ gates being decomposed. Note that the second decomposed CZ gate only doubles the space consumption of part A, because after this CZ gate there is no gate in part B performed on qubit $2$, thus the implicit decomposition.}
    \label{circuit_ex2}
\end{figure*}

For a better understanding of Technique 1, let us first consider a circuit example shown in Fig. \ref{circuit_ex2}.
The partition scheme for simulation is illustrated by the blue dotted line and yellow dotted line. The dotted lines cut two CZ gates, and partition the circuit into 3 parts: $A,B,C$. We use $|\phi\>$ and $|\xi\>$ to describe the initial states in the two subsystems. Because there are two CZ gates being decomposed (cut by dotted lines), four branching paths in total are generated. Let $|\phi^{out}_{l_1,l_2}\>$ be the resulting state after performing the gates in part A, $l_1$ and $l_2$ denote the indices of qubit $2$ in two different time. State $|\xi^{out}_{l_1,l_2}\>$ is similar to $|\phi^{out}_{l_1,l_2}\>$. Then we have\footnote{The order of performing gates in circuit is from left to right, while the matrix-vector multiplication is from right to left.}:
\begin{align*}
|\phi^{out}_{l_1,l_2}\> = X^0 CZ^{0,1} &Y^1 CZ^{2,1}_{l_2} T^1CZ^{2,1}_{l_1} H^0 H^1 |\phi\>
\end{align*}
where $l_1$ and $l_2$ denote the indices of qubit 2 when decomposing the cutting CZ gates, thus $CZ^{2,1}_{0} = I^1$ and $CZ^{2,1}_{1} = Z^1$. Furthermore, we have: 
$$
|\xi^{out}_{l_1,l_2}\> = X^3 P^2_{l_2} T^2 CZ^{2,3} P^2_{l_1} H^3 H^4 |\xi\>
$$
where $P$ is a projection operator: $P_i|i\> = |i\>$ and $P_i|1-i\> = 0$ for $i=0,1$. The starting state of part C is:
\begin{align}
|\psi^{in}\> = \sum_{l_1,l_2} |\phi^{out}_{l_1,l_2}\> \otimes |\xi^{out}_{l_1,l_2}\>
\end{align}
We perform the remaining gates in part C, and the eventually state $|\psi^{out}\>$ is:
\begin{align} \label{psi_out}
|\psi^{out}\> = Y^1 X^2 CZ^{1,2} X^2 |\psi^{in}\>
\end{align}
There are $2^4=16$ amplitudes to calculate for $|\psi^{in}\>$ (or $|\psi^{out}\>$). But we do not need memory space for 16 amplitudes(not counting memory space for $|\phi\>$ and $|\xi\>$) to accomplish the calculation. Note that in part C there are gates only performed on qubit 1 and qubit 2, and no gate on qubit 0 and qubit 3. So $|\psi^{in}\>$ can be divided into 4 blocks by enumerating the indices of qubit 0 and qubit 3. Let $|\psi^{q_1,q_2}_{i_0,i_3}\>$ denote the reduced state of qubit 1 and qubit 2 with qubit 0 at $|i_0\>$ and qubit 3 at $|i_3\>$. Then $|\psi^{in}\> = \bigoplus_{i_0,i_3}|\psi^{q_1,q_2}_{i_0,i_3}\>$. Equation (\ref{psi_out}) turns into:
\begin{align}\label{eq3}
    |\psi^{out}\> &= \bigoplus_{i_0,i_3}|\psi^{out,q_1,q_2}_{i_0,i_3}\> \nonumber\\
    &= \bigoplus_{i_0,i_3} (Y^1 X^2 CZ^{1,2} X^2 |\psi^{q_1,q_2}_{i_0,i_3}\>)
\end{align}
Now we know that if all the results of part A and B are calculated and stored in memory, there only needs extra space for $2^2 = 4$ amplitudes in part C. From now on we set an amplitude as a basic storage unit ($8+8=16\ bytes$), and the total space consumption is $S_A+S_B+4$ instead of $S_A+S_B+16$, where $S_A$ and $S_B$ are the space consumption of part A and B, respectively.

\par\setlength\parindent{1em} From the above analysis, we know $S_A = S_B = 2^{2+2} = 16$. However, $S_B$ can be halved without introducing extra computation. Note that after the second cut CZ (in red) gate being decomposed, there is no any gate performed on qubit $2$ in part B, so for any amplitude of $|\xi^{out}_{l_1,l_2}\>$, let it be $\xi_{i_2,i_3,l_1,l_2}$, where $i_2$ and $i_3$ are the indices of qubit 2 and qubit 3. We have
$$
\xi_{i_2,i_3,l_1,l_2 \neq i_2} = 0,\text { for}\ l_2=0,1
$$
Thus, the index $l_2$ could be absorbed into index $i_2$, and we only need to store $|\xi^{out}_{l_1}\>$. This means that when we finish the calculation of part B, only $2^{1+2}=8$ amplitudes to be stored, while in part A there are still $2^{2+2}=16$ amplitudes to be stored. Because the second cut CZ gate is decomposed but the space consumption need not be doubled for control part (part B in this example), we call this technique \textit{implicit decomposition}.
\par\setlength\parindent{1em}The implicit decomposition is useful when the space consumption of part A and B is not balanced, say there need space $S_A$ to store part A and some fewer space $S_B$ to store part B,  we could apply implicit decomposition to part A, and only making the space of part A to $S_A'$ while leaving the space of B unchanged, until $S_A'$ and $S_B$ are almost in the same magnitude.

\par
\begin{figure}
    \centering
    \includegraphics[scale=0.5]{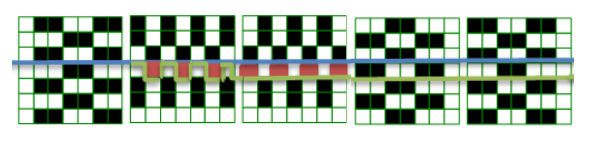}
    \caption{Implicit decomposition applied to 49-qubit universal random circuits. In this and following figures, two adjacent blocks represent a CZ gate. And the blocks in red represent the CZ gate that could be implicit decomposed. Because part A has 21 qubits and part B has 28 qubits, without implicit decomposition $S_B = 2^{28+7}$,$S_A=2^{21+7}$. After we use implicit decomposition on these seven cut CZ gates at the second and third layer, the ultimate space consumption of part B is $s'_B = S_B/2^{7}$, and now $S_A = S_B'$.}
    \label{circuit_ex3}
\end{figure}

\setlength\parindent{1em}For a universal random quantum circuit of $m\times n$ grid, the implicit decomposition works well when $m$ or $n$ is odd. Our main target is $7\times 7$-qubit circuits. At first, the two cutting lines are at the same position and partition the circuit into two parts: part A with 21 qubits, part B with 28 qubits. Then we apply this technique to part B, as shown in Fig. \ref{circuit_ex3}.

\subsubsection{Dynamic programming}

To reduce the total space consumption and make the simulator easier to optimize, we avoid decomposing CZ-gates between part C and parts A, B. At first two splitting line overlap. When the two splitting lines separate, they start to walk around the CZ gates and will not cut-off any one of them. A feasible partition scheme also requires the total space consumption less than the memory space. To find an optimal partition scheme under these constrains, we design a dynamic algorithm for state compression.

\par\setlength\parindent{1em} Let $f(t,i_1,i_2,i_3,i_4,i_5,i_6,i_7)$ denote the minimal space consumption (exponential in $f$) of part A when the partition scheme reaches layer $t$ and the position of upper splitting line is $(i_1,i_2,...,i_7)$. Here, $(i_1,i_2,...,i_7)$ is used to indicate the distance between the current position and the initial position. Initially, the upper splitting line is beneath the third row of qubits, and $f(1,0,...,0) = 21$. Note that $f(7,0,0,...,0) = 21+3 = 24$ and $f(8,0,0,...,0) = 21+3+4 = 28$. Since we have a restriction that no CZ gate between part A and B could be decomposed, $f(t,i_1,i_2,...,i_7)$ will be \textit{illegal} when $(i_1,i_2,...,i_7)\neq(0,0,...,0)$ and $(i_1,i_2,...,i_7)$ splits some CZ gate at layer $t$.

\begin{figure}
    \centering
    \includegraphics[scale=0.3]{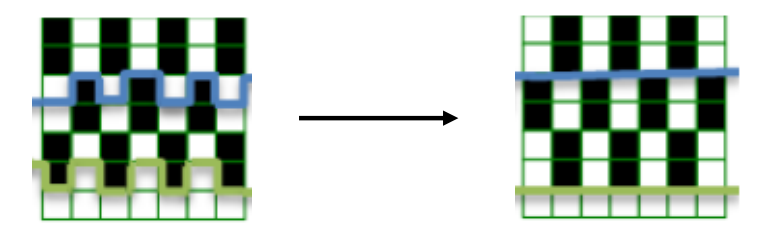}
    \caption{A illustrative example of legal transition between two adjacent layers. From layer $t$ to $t+1$, the position of upper splitting line (in blue) transforms from $(0,1,0,1,0,1,0)$ to $(1,1,...,1)$. Since $(1,1,...,1)$ does not cut any CZ gate at layer $t+1$, this is a legal transition. So $f(t+1,1,1,...,1) = min\{f(t+1,1,1,...,1),f(t,0,1,0,1,0,1,0)\}$.}
    \label{circuit_ex4}
\end{figure}

\begin{algorithm}
\caption{Pseudocode of the Dynamic Programming}
\KwIn{$f(t)$}
\KwOut{$f(t+1)$}
\For {$0\leq i_1,i_2,...,i_7 \leq 3$}{
    $f(t+1,i_1,i_2,...,i_7) = \infty$\;
    \If{position $(i_1,i_2,...,i_7)$ is not illegal}{\textbf{continue}\;}
    $cutnum = number\ of\ CZ\ gates\ splitting\ by\ (i_1,...,i_7)$\;
    \For{$0\leq j_1\leq i_1,0\leq j_2\leq i_2,...,0\leq j_7\leq i_7$}{
        $f(t+1,i_1,i_2,...,i_7) = min\{f(t,j_1,j_2,...,j_7)+cutnum,f(t+1,i_1,i_2,...,i_7)\}$\;
    }
}
\end{algorithm}
Similarly, let $g(t,i_1,...,i_7)$ denote the target function of part B. Then $g(1,0,0,...,0) = 28$ and $g(8,0,0,...,0) = 35$. The recursion from $g(t-1)$ to $g(t)$ is similar to $f$, and the only difference is that when $(i_1,i_2,...,i_7)$ first leaves the initial position it has a chance of applying implicit decomposition.

\par
\begin{figure}
    \centering
    \includegraphics[scale = 0.5]{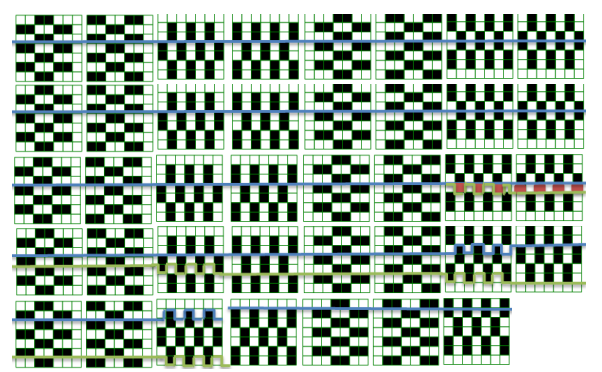}
    \caption{Partition scheme for 49-qubit circuit of depth 39. The scheme for depth 35 is exactly the front 35 layers of this figure, except for that at layer 35 two splitting lines are still straight as in layer 34 and cut 6 extra CZ gates (See the last layer of partition scheme for 49-circuit of depth 27 in Fig. \ref{fig:49qubit_depth27}). Note that the CZ gates at the last layer would not impact the probabilities and could be removed because they diagonal. Thus the number of cut CZ gates being cut (the implicit decomposed CZ gates and CZ gates at the last layer not included) is 14 for both depth 35 and 39, but in the case of depth 35 $S_C = 2^{28}$ and in the case of depth 39 $S_C = 2^{42}$.}
    \label{fig:49qubit_depth39}
\end{figure}

\begin{figure}
    \centering
    \includegraphics[scale = 0.35]{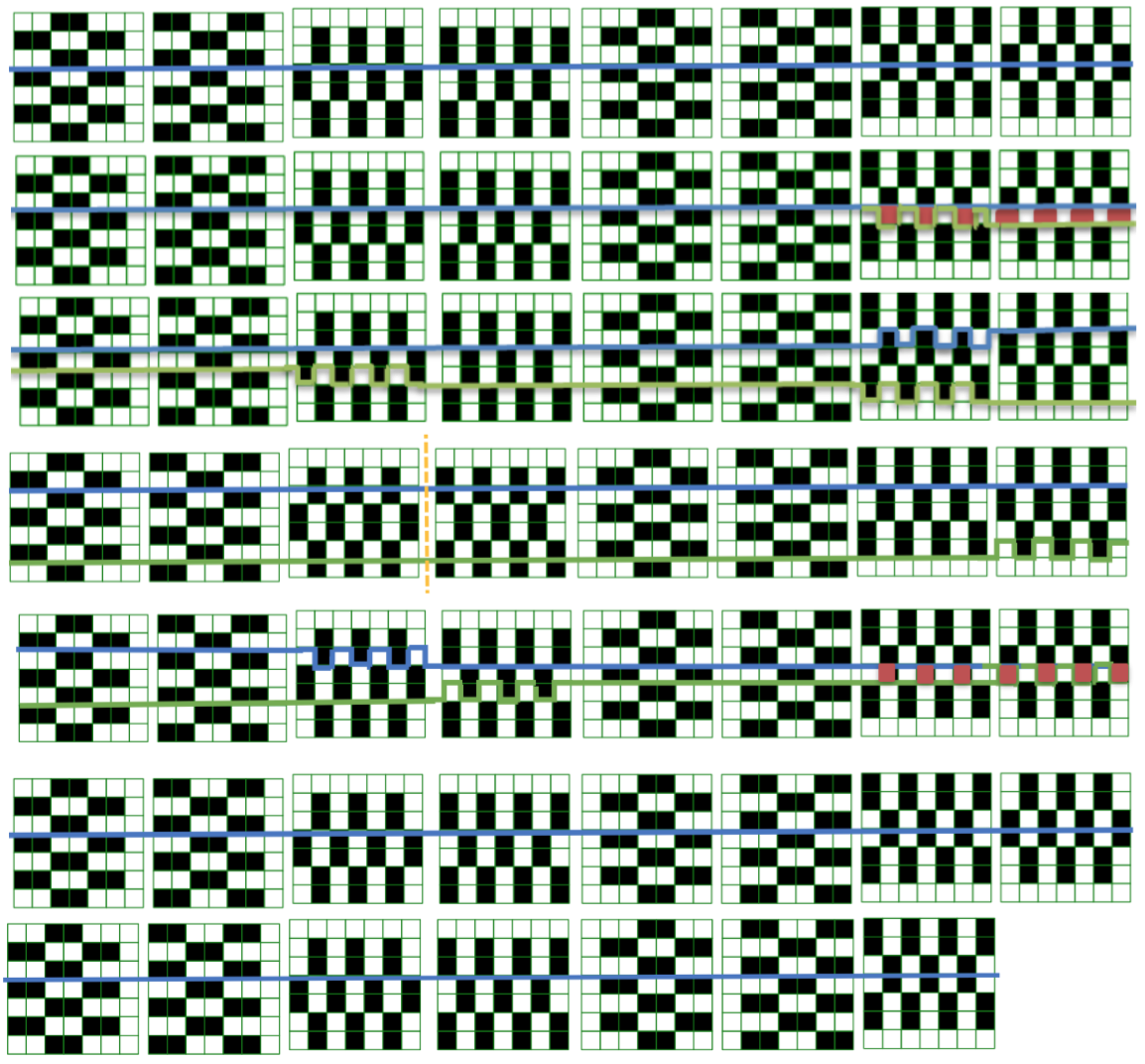}
    \caption{Partition scheme for calculating one amplitude for 49-qubit circuit of depth 55.}
    \label{fig:49qubit_depth27}
\end{figure}

\setlength\parindent{1em} To find a good partition scheme for circuit of depth $t$, we can traverse $f(t)$ and $g(t)$ to find an optimal combination of $f(t,i_1,i_2,...,i_7)$ and $g(t,j_1,j_2,...,j_7)$ which minimizes $S_A+S_B+S_C$, where 
$S_A = 2^{f(t,i_1,i_2,...,i_7)}$, $S_B = 2^{g(t,j_1,j_2,...,j_7)}$ and 
$S_C = 2^{\sum_{1\leq k\leq7}({i_k+j_k})}.$ 
Several partition schemes are illustrated in Fig. \ref{fig:49qubit_depth39} and Fig. \ref{fig:49qubit_depth27}. There need space $S_A = S_B  = 2^{35}$, $S_C = 2^{28}$ to execute the simulation for depth 27,  while $S_A = S_B = 2^{42}$  for depth 35 and 39. The maximal number of qubits that could be simulated on one node is 28. This indicates that from depth 35 to depth 39, there will be a drop in performance, which is shown in section \ref{sec33}.

\begin{table}[]
    \centering
    \begin{tabular}{|c|c|c|c|}
    \hline
       Depth & 27 & 35 & 39  \\
    \hline
       $S_A$ & $2^{35}$ & $2^{42}$ & $2^{42}$\\
    \hline
       $S_B$ & $2^{35}$ & $2^{42}$ & $2^{42}$\\
    \hline
       $S_C$ & $2^{28}$ & $2^{28}$ & $2^{42}$\\
    \hline
    \end{tabular}
    \caption{The space consumption of parts A,B,C for 49-qubit circuit of different depth.}
    \label{tab:my_label}
\end{table}

\subsubsection{Summary}
The two techniques proposed above not only work for universal random circuit. For circuit of an arbitrary 2-D grid structure, our method can find a proper partition scheme to reduce the time and space complexity of simulation. Table \ref{algorithmic comparison} gives an algorithmic comparison of our partition scheme and \cite{ibm}. 
\begin{table}[]
    \centering
    \begin{tabular}{|c|c|c|c|}
    \hline
    Reference & Depth & Space & Computation amount \\ 
    \hline
    \hline
    IBM\cite{ibm} & 27 & 64 TB & $2^{49}(2^{10}+n_{C})$ \\
    \hline
    Sunway & 27 & 1 TB & $2^{49}(2^{7}+n_{C})$ \\
    \hline
    \hline
    IBM\cite{ibm} & 39 & N/A & N/A \\
    \hline
    Sunway & 39 & 256 TB & $2^{49}(2^{14}+n_{C})$ \\
    \hline
    \end{tabular}
    \caption{An algorithmic comparison of our partition scheme and the partition scheme in \cite{ibm} for 49-qubit circuits. Space means the least memory needed to execute the simulation. $n_{C}$ is the number of gates in part C, usually several hundred. The computation amount means the number of float-point operations. No partition scheme for depth 39 is given in \cite{ibm}. This table shows that the combination of techniques 1 and 2 produces a more efficient algorithm.}
    \label{algorithmic comparison}
\end{table}

\par\setlength\parindent{1em}Obviously, $7\times 7$-qubit circuits of depth 39 and of depth 40 have different difficulties in physical preparation and operation. We provide an evidence showing that our method might reach depth 40 if the target circuit is a \textquotedblleft lucky circuit\textquotedblright\ (i.e., with enough T gates in special positions at layer 40). For example, using the tensor slice technique in \cite{ibm}, if there is at least one T gate in the four single-qubit gates on qubit $0,2,4,6$ at layer 40, the size of a slice is at most $2^{45}$, which could be directly simulated on Sunway, though the performance will further drop from depth 39 to depth 40. Because $X^{1/2},Y^{1/2},T$ appears randomly at the positions for single-qubit gates, the probability that a 49-qubit and 40-depth universal random circuit could be simulated on Sunway is:
$$
p = 1-(2/3)^4 = 65/81
$$

\begin{figure}
    \centering
    \includegraphics[scale = 0.8]{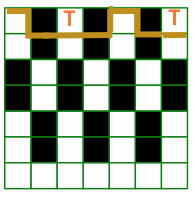}
    \caption{Tensor slicing technique in \cite{ibm}. This is a example that two T gates appears in the four positions, thus 5 qubits in total could be sliced.}
    \label{circuit_ex5}
\end{figure}

\subsection{Calculating one or a few amplitudes (Task 2)}

\begin{figure}
    \centering
    \includegraphics[scale = 0.3]{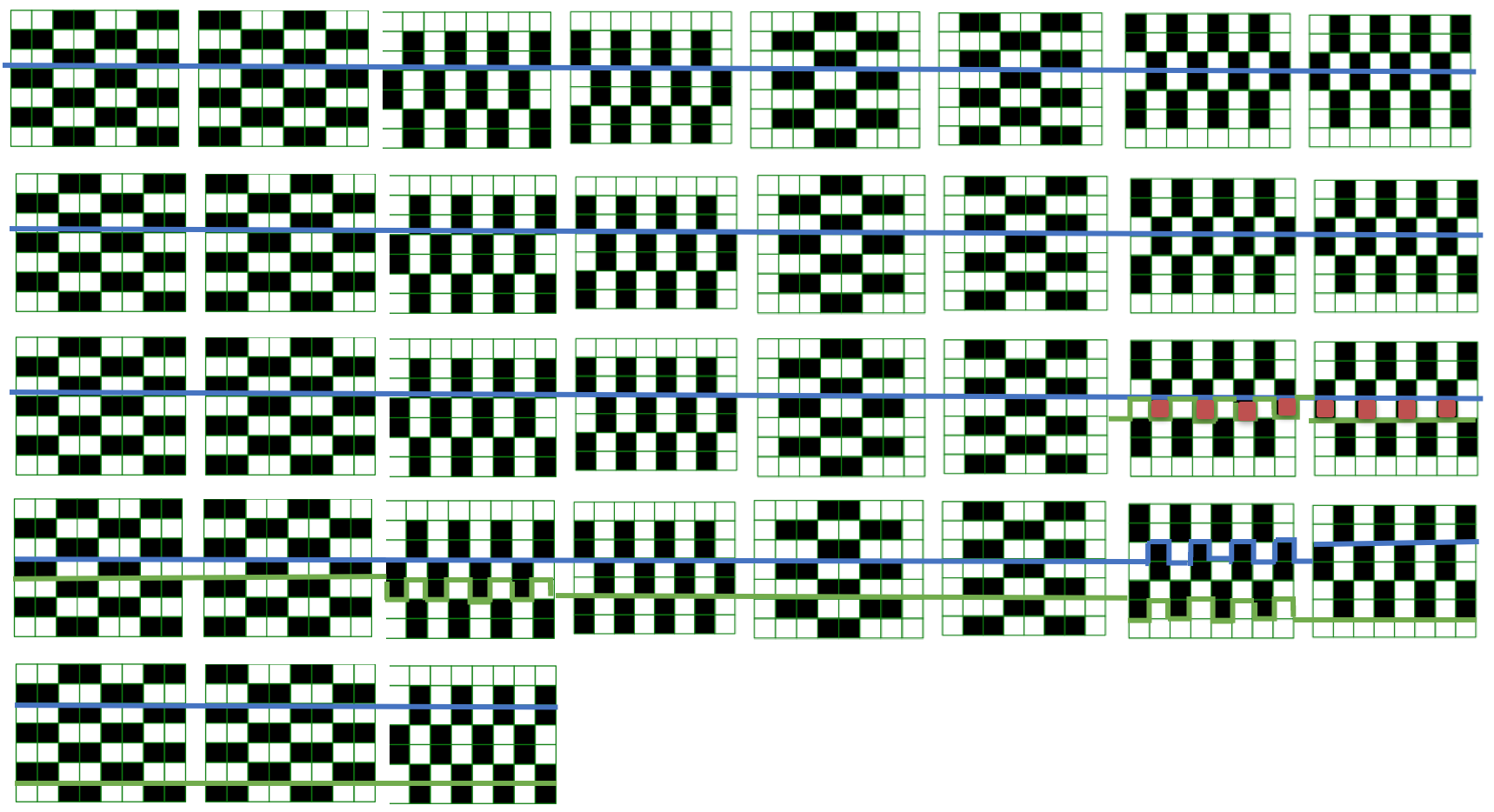}
    \caption{Partition scheme for 56-qubit circuits of depth 35. The method for solving a part of amplitudes from this circuit is exactly the same as computing a complete state-vector for 49-qubit circuits. Note that $S_A = S_B = 2^{48}$ here, which already exceed the memory limit of Sunway. Thus a little space-time tradeoff \cite{chenlijie} is needed. The space-time tradeoff, which is also needed for 64-qubit circuits with depth 30 and 72-qubit circuits with depth 27, can be achieved by simply enumerating the first several decomposed CZ gates \cite{ustc}.}
    \label{fig:56-qubit}
\end{figure}

\par\setlength\parindent{1em}The method in Section \ref{sec21} can be used to compute the complete state-vector for 49-qubit circuits. For circuits of 56 qubits or larger size, it is difficult to calculate all the amplitudes due to limited time and space. However, to test the fidelity of a real quantum circuit, one only needs to sample (i.e. calculate a small number of) amplitudes, usually ranging from $10^3$ to $10^6$\cite{googlequantumsupremacy}. Our method can finish this task very efficiently because all the amplitudes of eventually states in part A and B are stored in memory. For example, in the cases of $7\times8$ qubits with depth 35, or $8\times8$ qubits with depth 30 is easy to calculate a large amount of (e.g. $\geq 2^{32}$) amplitudes (in less than 1 hour). Figure \ref{fig:56-qubit} shows the partition scheme for 56-qubit circuits with depth 35. The schemes for 64-qubit and 72-qubit circuit are similar.

\par\setlength\parindent{1em}When focusing on a $7\times 7$-qubit circuit, we will introduce a special and straight method to calculate an amplitude of the final state. The target is to calculate $$\alpha_x = \<x|\mathcal{U}_{circuit}H^{\otimes 49}|00...0\>$$ for $0\leq x \leq 2^{49}-1$. Since we could calculate the complete state-vector for a $7\times 7$-qubit circuit of depth 27, we can also sample one amplitude for a circuit of depth 55. Let ${U}_{circuit} = \mathcal{U}_2 \mathcal{U}_1$ in which $\mathcal{U}_1$ has 27 layers and  $\mathcal{U}_2$ has 28 layers, $|\psi\> = \mathcal{U}_1H^{\otimes 49}|00...0\>$ and $|\varphi\> = \mathcal{U}_2^{\dagger}|x\>$, we have:
$$
\alpha_x = \<\varphi|\psi\>
$$
Thus, we calculate $|\varphi\>$ and $|\psi\>$  simultaneously in memory, during the calculation we computing the inner product of every two corresponding blocks of $2^{28}$ amplitudes of $|\varphi\>$ and $|\psi\>$. Sum all $2^{21}$ inner products and we get $\alpha_x$. Because the least space consumption of simulating a circuit of depth 27 is O($2^{35}$), this method can be parallelized to calculate more amplitudes.

\subsection{Optimization for reducing communication amount} \label{sec_eq6}
The method presented above concerns mainly the memory space limitation of Sunway. But if the network communication amount is too large in implementing this method, it will be impractical since the network bandwidth of Sunway is limited. In this subsection we describe a method to reduce the communications of a key step in our simulation. We will first explain why the communication is needed and then describe how to optimize it.

Recall that equation (\ref{eq3}) is the final step to compute the complete state-vector, and this step can be divided into $2^{n}/S_C$ subtasks which could be paralellized, where $n$ is the number of qubits in the whole circuit and $S_C$ is the space consumption of part C. For 49-qubit circuits of depth 27 and depth 35, (\ref{eq3}) can be rewritten as:
\begin{align}\label{eq4}
    &|\psi^{out}\> = \bigoplus_{i_0,i_1,...,i_{13},i_{42},i_{43},...,i_{48}} |\psi^{out,q_{14},q_{15},...,q_{41}}_{i_0,i_1,...,i_{13},i_{42},i_{43},...,i_{48}}\> \nonumber\\
    &= \bigoplus_{i_0,i_1,...,i_{13},i_{42},i_{43},...,i_{48}} (\mathcal{U}_C|\psi^{q_{14},q_{15},...,q_{41}}_{i_0,i_1,...,i_{13},i_{42},i_{43},...,i_{48}}\>)
\end{align}
where $\mathcal{U}_C$ denotes the unitary operation represented by part C. Note that calculating $|\psi^{out,q_{14},q_{15},...,q_{41}}_{i_0,i_1,...,i_{13},i_{42},i_{43},...,i_{48}}\>$ in equation (\ref{eq4}) is essentially the same as simulating a 28-qubit circuit, which could be finished in a single node. Now the remaining problem is to efficiently prepare $|\psi^{q_{14},q_{15},...,q_{41}}_{i_0,i_1,...,i_{13},i_{42},i_{43},...,i_{48}}\>$ for each set of possible values of $(i_0,i_1,...,i_{13},i_{42},i_{43},...,i_{48})$, where $i_k \in\{ 0,1\}$. Note that
\begin{align} \label{eq5}
   & |\psi^{q_{14},q_{15},...,q_{41}}_{i_0,i_1,...,i_{13},i_{42},i_{43},...,i_{48}}\> = \nonumber \sum_{l_1,l_2,...,l_t,i_{21},i_{22},...,i_{27}}\\ &\ \ \ \ \ \ \ \ \ \ (|\phi^{out,q_{14},q_{15},...,q_{20}}_{i_0,i_1,...,i_13,l_1,l_2,...,l_t,i_{21},i_{22},...,i_{27}}\>\otimes \nonumber\\ & |\xi^{out,q_{21},q_{22},...,q_{42}}_{i_{42},i_{43},...,i_{48},l_1,l_2,...,l_t,i_{21},i_{22},...,i_{27}}\>)
\end{align}
where $t$ is the number of decomposed CZ gates, $t=7$ for depth 27, and $t=14$ for depth 35. $|\phi^{out,q_{14},q_{15},...,q_{20}}_{i_0,i_1,...,i_6,l_1,l_2,...,l_t,i_{21},i_{22},...,i_{27}}\>$ is a $2^{7}$-length state-vector, where the indices of qubit 0-6 are $i_0,i_1,...,i_6$, the indices of control qubits of $t$ decomposed CZ gates are $l_1,l_2,...,l_t$, and the indices of control qubits of 7 implicit decomposed CZ gate are $i_{22},...,i_{27}$. Similarly, $|\xi^{out,q_{21},q_{22},...,q_{41}}_{i_{42},i_{43},...,i_{48},l_1,l_2,...,l_t,i_{21},i_{22},...,i_{27}}\>$ is a $2^{21}$-length state-vector. Because of implicit decomposition, for each value set of $i_{21},i_{22},...,i_{27}$, $|\xi^{out,q_{21},q_{22},...,q_{42}}_{i_{42},i_{43},...,i_{48},l_1,l_2,...,l_t,i_{21},i_{22},...,i_{27}}\>$ has only $2^{14}$ non-zero amplitudes. Thus, (\ref{eq5}) can be rewritten as:
\begin{align} \label{eq6}
   & |\psi^{q_{14},q_{15},...,q_{41}}_{i_0,i_1,...,i_{13},i_{42},i_{43},...,i_{48}}\> = \sum_{l_1,l_2,...,l_t} (\bigoplus_{i_{21},i_{22},...,i_{27}} \nonumber\\ &
    |\phi^{out,q_{14},q_{15},...,q_{20}}_{i_0,i_1,...,i_6,l_1,l_2,...,l_t,i_{21},i_{22},...,i_{27}}\> \otimes \nonumber\\ & |\xi^{out,q_{28},q_{29},...,q_{42}}_{i_{42},...,i_{48},l_1,...,l_t,q_{21}=i_{21},...,q_{27}=i_{27}}\>)
\end{align}
where $|\phi^{out,q_{14},q_{15},...,q_{20}}_{i_0,i_1,...,i_6,l_1,l_2,...,l_t,i_{21},i_{22},...,i_{27}}\>$($|\phi\>$ for short) is still of $2^{7}$-length but $|\xi^{out,q_{28},q_{29},...,q_{42}}_{i_{42},...,i_{48},l_1,...,l_t,q_{21}=i_{21},...,q_{27}=i_{27}}\>$($|\xi\>$ for short) is of $2^{14}$-length.

Now we consider how to realize equation (\ref{eq6}). The data from part A have $2^{t+7}$ complex numbers (amplitudes), and the data from part B have $2^{t+14}$ complex numbers. Thus, for the case of depth 35, the data from part B have $2^{14+21} = 2^{35}$ complex numbers. Directly calculating (\ref{eq6}) needs $2^{7}$ nodes to communicate (e.g. an \emph{MPI\_Gather} is feasible). This is very inefficient, because there are $2^{49-28} = 2,097,152$ entities of (\ref{eq6}) to calculate, and each entity needs an \emph{MPI\_Gather} in 128 nodes. Assume that we have $2^{15} = 32,768$ free nodes for this work. One round can proceed $2^{15-7} = 256$ \emph{MPI\_Gather}s and calculate the same amount of entities. Then we need $\frac{2,097,152}{256} = 8192$ rounds of \emph{MPI\_Gather} to finish all the jobs.

However, we can slightly change the form of (\ref{eq6}) to:
\begin{align}\label{eq7}
   & |\psi^{q_{14},q_{15},...,q_{41}}_{i_0,i_1,...,i_{13},i_{42},i_{43},...,i_{48}}\> = 
    \bigoplus_{i_{21},i_{22},...,i_{27}}(\sum_{l_1,l_2,...,l_t}\nonumber\\ &
    |\phi^{out,q_{14},q_{15},...,q_{20}}_{i_0,i_1,...,i_6,l_1,l_2,...,l_t,i_{21},i_{22},...,i_{27}}\>\otimes \nonumber\\ & |\xi^{out,q_{28},q_{29},...,q_{42}}_{i_{42},...,i_{48},l_1,...,l_t,q_{21}=i_{21},...,q_{27}=i_{27}}\>)
\end{align}
Note that for each element \textit{in the bracket}, the data from part A and B have $2^{21}$ and $2^{28}$ complex numbers, respectively. This can be finished in a single node and the length of result is $2^{21}$. We further append $|\phi\>$ with qubit 7-13 so that $|\phi\>$ also becomes a $2^{14}$-length vector:
\begin{align*}
    |\phi\> = |\phi^{out,q_{7},q_{9},...,q_{20}}_{i_0,i_1,...,i_6,l_1,l_2,...,l_t,i_{21},i_{22},...,i_{27}}\>
\end{align*}
Now $|\phi\>\otimes|\xi\>$ is of $2^{28}$-length and can still be calculated in a single node. Actually, it can be regarded as a 28-qubit state:$|\psi^{q_7,...,q_{20},q_{28},...,q_{42}}\>$. Our target is $|\psi^{q_{14},...,q_{27},q_{28},...,q_{42}}\>$. Note that for every group of 128 nodes, they store $|\psi^{q_8,...,q_{20},q_{28},...,q_{42}}\>$ for $(i_{21},i_{22},...,i_{28})=(0,0,...,0)$ to $(1,1,...,1)$. Thus, by performing an \emph{MPI\_Alltoall} on these 128 nodes, we get 128 entities of (\ref{eq6}). Assume again that we have $2^{15}$ free nodes to work for part C. Then  we can get $2^{15}$ entities in a single round. Furthermore, in only $2^{21-15} = 64$ rounds, the whole task can be finished.

Solving (\ref{eq6}) for 49-qubit circuit of depth 39 is essentially the same as the case of depth 35. Table \ref{com. opt.} shows the improvement in network communication that the optimization brings.

\begin{table}[]
    \centering
    \begin{tabular}{|c|c|c|c|}
    \hline
    Optimization & Main work & Overall time & Speedup \\
    \hline
    without & $2^{21}$ MPI\_Gathers & 1021.9 min & 1\\
    \hline
    with & $2^{14}$ \emph{MPI\_Alltoall} & 17.8 min & 57.4\\
    \hline
    \end{tabular}
    \caption{The comparison of network communication amounts for computing Eq. (\ref{eq6}) with and without optimization. Every single MPI\_Gather or {MPI\_Alltoall} is within 128 nodes, so they can be executed in parallel. The overall time is under the condition of using 32768 nodes for network communication. For network communication without optimization the overall time is only an estimation since we only executed $2^{15}$ MPI\_Gathers, and multiplies by 64 the execution time, which is 15.96 minutes.}
    \label{com. opt.}
\end{table}

\subsection{Single node optimizations}\label{section2.3}

In this subsection, we introduce our optimizations for the quantum circuit simulation at the single node (single core group) scale. The optimizations for every single node is very important for improving the performance of our method.
\par\setlength\parindent{1em} As agreed in Section \ref{section1}, one node represents a core group in Sunway, and it has 1 master core and 64 slave cores. Each node has 8GB DDR memory, shared by both master and slave cores. The maximum qubit number that could be simulated on one node is 28 if we use two doubles to represent an amplitude, since $2^{28}\times 16B = 4GB$.  To obtain full power of Sunway TaihuLight, one must distribute most of the computing tasks to slave cores. Each slave core has a private and separate memory unit called \emph{local data memory (LDM)} (also known as \emph{scratch pad memory}) of 64kB size. To execute high-speed calculations a slave core must fetch data from the main memory to its own LDM and keeps it in LDM for calculation as long as possible. This fetching-data behavior is usually called \emph{direct memory access} (DMA). To get high DMA bandwidth, it usually requires the data fetched consecutive.

\par\setlength\parindent{1em} If LDMs fetch data from the main memory for every gate performed, and put the data back to main memory when the calculation is finished, the simulation will be inefficient due to low flop-to-Byte ratio\footnote{A $2\times2$ complex matrix multiplying a $1\times2$ complex vector needs 14 float-point operations. For a 28-qubit circuit, each non-diagonal gate requires $2^{28-1}\times 14$ float-point operations and 4G DMA \emph{get} and \emph{put}}.  In our experiments, the average execution time is 0.32s per gate in such way, thus the performance is bounded by DMA speed\footnote{ The DMA \emph{get} bandwidth and \emph{put} bandwidth are both less than 25GB/s in our experiments}. However, we can take advantage of the data locality in a better way to reduce the DMA amount. For example, if each slave core has fetched from the main memory 16KB data in its LDM, that is, $2^{14-4} = 2^{10}$ amplitudes. For any gate performed on those qubits with their ranks lower than 10 (0 is the lowest rank), the calculation can be executed in this 16 KB data, i.e. $\alpha_i$ and $\alpha_{i+2^k}$ are in the same LDM. Thus we can perform a bunch of gates acting on low-rank qubits once instead of performing the circuit gate by gate. We call these 10 qubits with lowest ranks \emph{local qubits}, and other 18 qubits are \emph{global qubits}\footnote{this is different to \cite{massively2007,eth}. Their distinction of 'global' and 'local' is about the multi-node and single-node level. While our 'global' corresponds to main memory, and 'local' corresponds to LDM}. 

This idea is similar to the gate fusion techniques in \cite{qhipster}, which deals with the gates on low-rank qubits in cache. However, gate fusion is one-off, which can only be applied at the start of the circuit. 

To make the above procedure repeatable, we also adopt the qubit reordering method. Qubit reordering are used in \cite{massively2007} and \cite{eth} to reduce the network communications. Here, our aim is to maintain the data locality and reduce the amount of memory access. That is, only diagonal gates, or non-diagonal gates on local qubits are calculated. To accomplish this, we need to execute two types of qubit rank swaps:
\begin{itemize}
    \item Swap the qubits of rank 0-9 and qubits of rank 11-20
    \item Swap the qubits of rank 14-20 and qubits of rank 21-27
\end{itemize}
With a gate scheduling preprocessing program, which also utilizes the diagonal properties of gate $T$ and $CZ$, we get the amount of swaps for 28-qubit quantum supremacy circuit:
\begin{table}[]\label{swap_times}
    \centering
    \begin{tabular}{|c|c|c|c|c|}
        \hline
    Depth  & 25 & 30 & 34 & 38 \\
         \hline
    Swaps  & 6 & 7 & 8 & 9 \\
         \hline
    \end{tabular}
    \caption{Frequency of swaps for 28-qubit universal random circuits with different depth, in the case that the number of local qubits is 10.}
    \label{tab:my_label}
\end{table}
We achieve fast swaps of qubit rank with the help of slave cores. Swapping the qubits of rank 0-9 and qubits of rank 10-19 is essentially a transpose of a complex matrix. The dimension of this matrix is $2^{10}\times2^{10}$, and $2^{28-20} = 256$ matrices in total need to be transposed. Swapping the qubits of rank 14-20 and qubits of rank 20-27 is similar, which is equivalent to a transpose of a $2^8\times2^{8}$-dimensional matrix, but each element of this matrix contains $2^{12}$ amplitudes.

\begin{figure}
    \centering
    \includegraphics[scale = 0.25]{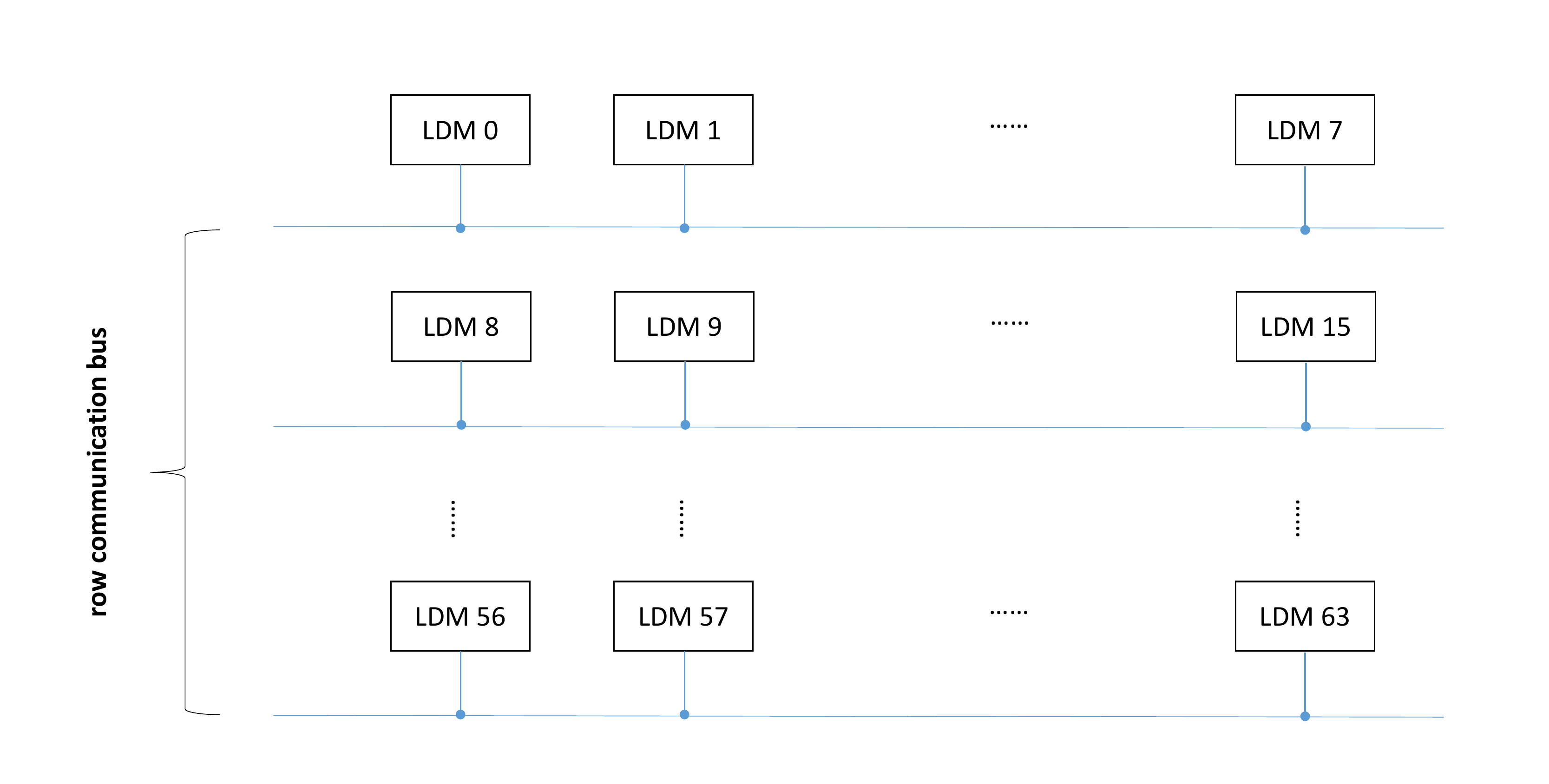}
    \caption{Register communication. There are 8 row communication buses and 8 column communication buses in a core group. In this figure only row communication buses are plotted, since in our experiments only row communication is used. In fact, the row communication is used to achieve the swap between qubits of rank 8-10 and qubits of rank 11-13. Register communication has very high bandwidth, more than 200GB/s in total for 8 row communication buses. So the cost of this step is very small, and we can treat 8 LDMs in a row as new a composite LDM. Thus the number of local qubits turns into 14.}
    \label{fig:my_label}
\end{figure}

\par\setlength\parindent{0em} \emph{Register communication} is a unique function in Sunway CPU, designed for fast data transmission between LDMs. The qubit reordering can be further optimized using this feature. Because the slave cores in a row can send/receive messages in a communication bus, we can treat a row of 8 separate LDMs as a composite LDM, while the data exchange between these 8 LDMs is fulfilled by register communication. If each slave core fetches 32kB consecutive data from the main memory to its LDM, there will be 11 local qubits. But when considering a composite LDM formed by a row of LDMs, the number of local qubits turns into 14. Thus we only need to execute one type of qubit swap:
\begin{itemize}
    \item Swap the qubits of rank 0-13 and qubits of rank 14-27
\end{itemize}
This is simply a transpose of a $2^{14}\times 2^{14}$-dimensional complex matrix, which can be quickly accomplished by slave cores.

Because of the high bandwidth of register communication, the time consumed on register communication is very small, so as to improving the overall performance of single node case.

\subsection{Other standard optimizations}
In this subsection we briefly introduce some other standard optimizations provided by Sunway TaihuLight, which can be exploited for our simulation of quantum circuits.
    \subsubsection{Vectorization}
        Sunway TaihuLight provides many 256-bit data types. In our simulation the type $doublev4$ is adopted for vectorization.
        The data stored in LDM is a complex number array with one double as the real part of a complex number and another double as its imaginary part. To vectorize the double-precise float-point calculation, we put four amplitudes into two $doublev4$ registers $v_r,v_i$ once. However, the real parts of these amplitudes are not consecutive, the imaginary parts neither. We use the instruction \emph{vshuffle} to solve this problem.
        
    \subsubsection{Instruction Reordering}
    Another optional optimization is instruction pipeline. The $put$ and $store$ operations, together with the multiply-add operations, can form a pipeline to further reduce the calculation time, especially when the data dependency between adjacent instructions is little. This optimization further improves the computational efficiency.

\section{Numerical Experiments and Results}

\subsection{Setup of Experiments}

Sunway TaihuLight is one of the most powerful supercomputer with over 100 Pflops computing capacity \cite{top500}. To test the limitation of our simulator on Sunway, we used 131072 nodes (32768 cpu chips), which is around $80\%$ of computing resource of the whole machine with nearly 1PB main memory in total. We implemented our simulator in C++ for master core managing programs and C for slave core computing programs. We use MPI for inter-node communications. To facilitate the most of computing capacity of Sunway we have used the \emph{athread} library \cite{sunway2016}.

According to previous analysis, a simulation task in our implementation has two stages:
\begin{itemize}
    \item stage 1: computing the results for part A and B and store them in the memory;
    \item stage 2: using the the results in stage 1 to generate the input and compute the results for part C, so as to solving all the amplitudes;
\end{itemize}
Stage 1 can be finished in 10 minutes if there is enough space to store the results for part A and B. Stage 2 thus is the bottleneck of the whole task. As illustrated in section \ref{sec_eq6}, stage 2 could be evenly divided into 64 rounds, making it convenient to parallelize and providing good strong scalability.

Stage 2 has two computing kernels: the first one is generating the input for part C, more precisely, computing the entities of eq(\ref{eq6}); the second one is simulating a 28-qubit circuit on single nodes. We call the first kernel \emph{tensor} because it calculates the tensor product of two complex vectors and sums them. We call the second kernel \emph{sim}.

\subsection{Performance Measurement}

The performance is usually computed in two ways:
\begin{itemize}
    \item Manually counting all double-precision arithmetic instructions in the assembly code;
    \item Using the hardware performance monitor of Sunway, PERF, to get the amount of double-precision arithmetic instructions retired on the CPE cluster.
\end{itemize}
Both ways provide similar results of counting the arithmetic operations. We employ the second way (PERF) in our study. And we obtain the sustained performance of two kernels: Kernel \emph{tensor} achieves 92.8 GFlops per core group; kernel \emph{sim} achieves 37.1 GFlops per core group. The performance of these two kernels fits the overall performance when simulating a 49-qubit circuit of depth 39.

\par\setlength\parindent{1em}Table \ref{28qubit} shows the performance of kernel \emph{sim} and speedup under cases of 28-qubit circuits with different depth. The number of global qubits is 14. Speedup is the speed-up ratio to the method of performing gate by gate without any optimization but using the slave cores to accelerate.

\begin{table}[]
    \centering
    \begin{tabular}{|c|c|c|c|c|c|}
    \hline
       Depth & 18 & 26 & 34 & 42 & 50 \\
    \hline
       Gates & 258 & 375 & 492 & 609 & 726\\
    \hline
       Swaps & 3 & 4 & 5 & 6 & 7\\
    \hline
       Time & 15.4s & 22.6s & 29.6s & 36.7s & 44.3s\\
    \hline
       Speedup & 7.04 & 6.95 & 6.96 & 6.95 & 6.88\\
    \hline
    \end{tabular}
    \caption{Performance of simulating 28-qubit circuits with different depth on a single node. The number of global qubits is 14. Speedup is the speed-up ratio to method of performing gate by gate without any optimization but using the slave cores to accelerate.}
    \label{28qubit}
\end{table}

\subsection{Time-to-solution} \label{sec33}

For the task of simulating a 49-qubit circuit of depth 35 and computing the complete state-vector, it takes around 3.7 hours.  The bottleneck is (\ref{eq6}), because solving $2^{21}$ entities of (\ref{eq6}) needs $2^{21+7+35} = 2^{63}$ times of complex number multiplication. This step occupied around 90\% of the run-time in the simulation of 35-depth circuit. For the task of simulating a 49-qubit circuit of depth 39, it takes around 4.2 hours. The reason for causing this drop in performance is that part C has $42$ qubits in the case of depth 39, while part C only has $28$ qubits in the case of depth 35. Thus in the case of depth 35, the calculation in part C are all within single nodes. While calculating a block of part C in the case of depth 39 is essentially simulating a 42-qubit circuit of depth 15, which needs one all-to-all communication on $2^{14}$ nodes \cite{eth}. Because there are $2^{49-42} = 128$ blocks to calculate, the amount of communication increases a lot.

The sustained performance is 4.92 PFlops for the case of depth 35 and 4.3 PFlops for the case of depth 39. 4.3 PFlops is around $3.44\%$ of the peak performance of Sunway. There are two reasons for this low efficiency: 1) we only use 131072 nodes, which is just around $80\%$ of the whole computing resource of Sunway; 2) for the cases of depth 35 and depth 39, the kernel \emph{tensor} is the bottleneck. However, during the simulation only half of the nodes are used for kernel \emph{tensor} due to the limited memory space of each node\footnote{In our implementation at current stage, 1/4 of the nodes need to store the results for part A, another 1/4 of the nodes need to store the entities of eq(\ref{eq6}), they can not participate in the computation of kernel \emph{tensor}.}. If we can find a better method for memory allocation we might let all 131072 nodes work for kernel \emph{tensor}, doubling the overall performance roughly.

\subsection{Improvement over previous works}

To show the improvement that our method brings, we first compare the overall performance between our work and the IBM's work \cite{ibm}, in the case of 49 qubits with depth 27. In principle, simulating the circuit with such a depth only needs 1024 nodes using our method. Increasing the nodes will decrease the time-to-solution. For a fair comparison, the performance of machine should be taken into consideration. We choose the result using 16384 node for comparison. It is $10.24\%$ of Sunway. And the improvement is shown in Table \ref{compare1}. The main reason for such improvement is also given.

\begin{table*}[]
    \centering
    \begin{tabular}{|c|c|c|c|c|c|}
    \hline
    Work & Qubits & Depth & Rmax (TFlop/s) & Time-to-solution & Speedup \\
    \hline
    IBM & 49 & 27 & 17,173.2 & $\geq 24$ hrs & 1\\
    \hline
    Sunway & 49 & 27 & 9,524.7 (16384 nodes) & 1.49 hrs & $\geq 29$ \\
    \hline
    \end{tabular}
    \caption{The performance comparison between our work and IBM simulator \cite{ibm}. \emph{RMAX} means the maximal sustained performance \cite{top500}. $10.24\%$ of Sunway has the $Rmax$ of $9,524.7$ TFlop/s. The speedup is calculated by: $speedup = \frac{IBM\_time \times IBM\_Rmax}{Sunway\_time \times Sunway\_Rmax}$. The main reason for such speedup is that in \cite{ibm}, 10 CZ gates are decomposed in the case of 49 qubits with depth 27. While in our method there are only 7 decomposed CZ gates benefiting from implicit decomposition and dynamic programming techniques.}
    \label{compare1}
\end{table*}

\begin{table*}[]
    \centering
    \begin{tabular}{|c|c|c|c|c|c|c|c|c|}
    \hline
    Platform & Local qubits & Depth & Gates & Time-to-solution & Time per gate & Memory access per gate & Single node bandwidth & Speedup \\
    \hline
    Cori \uppercase\expandafter{\romannumeral2} & 30 & 25 & 369 & 9.58 s & 0.026 s & 16 GB & 460 GB/s & 1\\
    \hline
    Sunway & 28 & 26 & 375 & 22.6 s &  0.060 s & 4 GB & 27 GB/s & 1.84\\
    \hline
    \end{tabular}
    \caption{Analysis of two highly optimized simulator in the single node cases. We compare their average performance per unit memory bandwidth (say 1 GB/s). We reemphasize that in this article, one single node only means one core group of Sunway. While a SW26010 cpu chip has 4 core groups, its memory bandwidth quadruples, which is more than 100GB/s. Counting in the memory access and average time per gate, we get an approximate speedup: $sppedup = \frac{(ETH\_time\_per\_gate \times ETH\_MA\_per\_gate)/ Cori\_single\_node\_bandwidth}{(Sunway\_time\_per\_gate \times Sunway\_MA\_per\_gate)/Sunway\_single\_node\_bandwidth} = 1.84$}
    \label{compare2}
\end{table*}

\par\setlength\parindent{1em}Another reason for the speedup in Table \ref{compare1} is our single node optimizations, which make better use of the machine performance. To make this claim more convincing, we compare our single node optimizations with the ETH's work \cite{eth}, in which their single node case is highly optimized too. Again, to make comparison fair, we should consider a rather similar benchmark. Because the bottleneck of quantum circuit simulation in single node case is \textit{memory access}, we consider the memory bandwidth of one node in either machine. See Table \ref{compare2}.

The comparison with the ETH's work is not absolutely fair, but at least demonstrates that our single node optimizations are also very efficient to deal with the large amount of memory access even in nodes without high memory bandwidth.

\subsection{Scalability}

\begin{figure}
    \centering
    \includegraphics[scale=0.4]{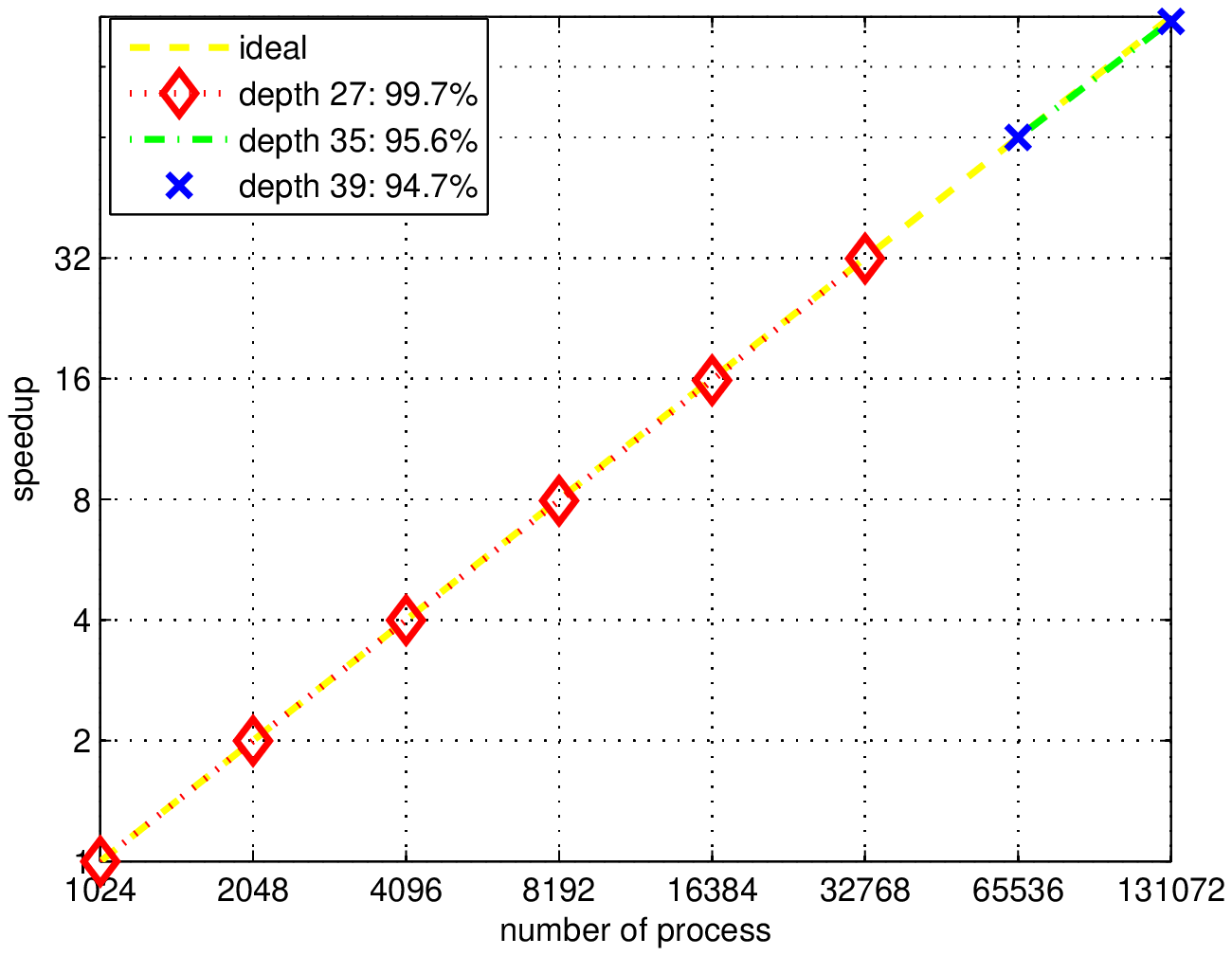}
    \caption{Strong scaling behaviour for the cases of depth 27, 35 and 39. The parallel efficiency under these three cases is also illustrated in the figure.}
    \label{fig:scaling27}
\end{figure}

 Figure \ref{fig:scaling27} shows the strong scaling behavior for circuits of different depth. As the stage 1 of computing results for part A and B usually takes a few minutes, this causes the drop of parallel efficiency especially when each node executes few rounds of stage 2. Note that the time consumption of stage 1 for the case of depth 27 is much less than that for the case of depth 35 and depth 39, so the parallel efficiency for the case of depth 27 is slightly better the other two cases. Moreover, simulating a circuit of depth 35 or depth 39 requires at least 65536 nodes.

\begin{figure}
\centering
    \includegraphics[scale=0.25]{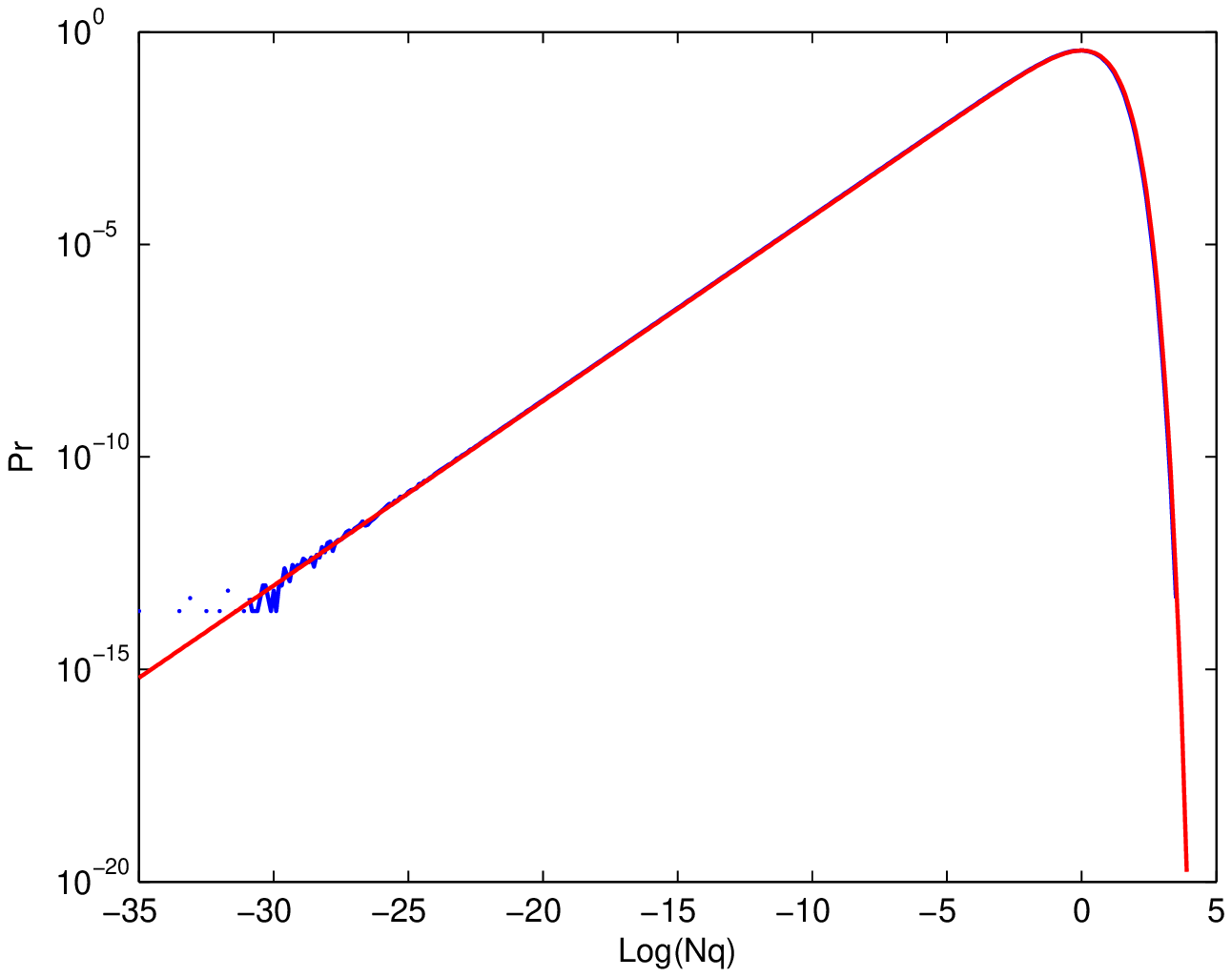}
    \includegraphics[scale=0.25]{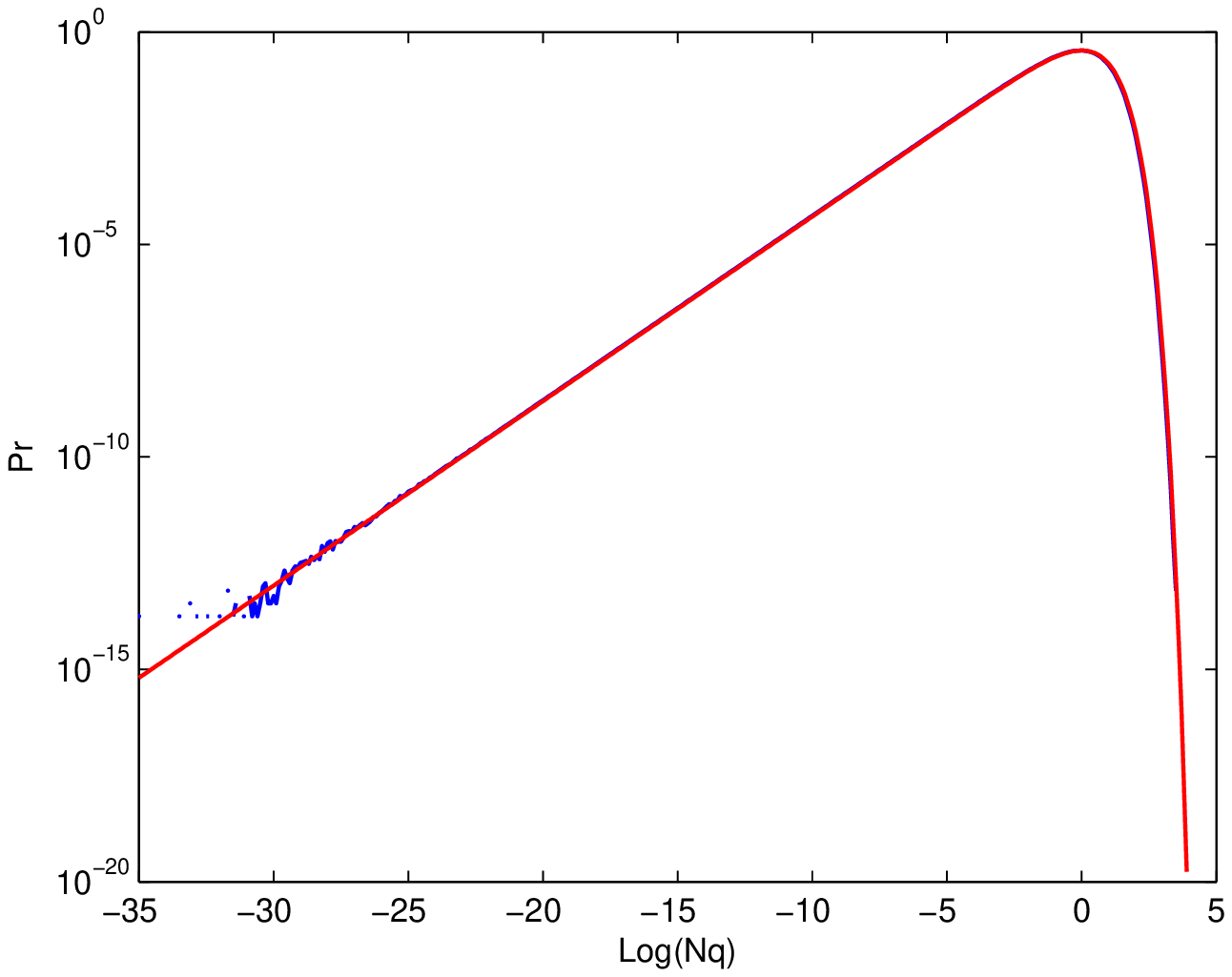}
    \caption{Histograms of log-transformed outcome probabilities for 49-qubit circuits, compared to theoretical Porter-Thomas distribution \cite{googlequantumsupremacy}. The left is the result of simulating a circuit of depth 35. The right is the result of circuit of depth 39. Red lines mean the theoretical Porter-Thomas distribution, and blue lines respresent the distribution of our experimental results. Both results fit the theoretical distribution well.}
    \label{fig:log_transform}
\end{figure}

\section{Conclusion and Future works}
This paper describes our method and implementation of quantum circuit simulator on Sunway TaihuLight. The results indicate that for current universal random circuits, 49 qubits with depth 39 is reachable. To find a proper bound of quantum supremacy in terms of universal random circuits, one might 1) increase the depth or qubits of the circuits; or 2) modify the structure of current universal random circuits. Whatever, classical computers have their limits on simulating quantum circuits. We believe there will be one day that quantum computer can solve certain problems which classical computers cannot. Before that day comes, simulating quantum circuits on classical computers is crucial to understand the power and limit of quantum computers. Even after that day, a simulator of quantum circuits on a classical computer will still be helpful for design, synthesis, testing and verification of quantum circuits.

\par\setlength\parindent{1em}Follow-up work of this paper includes further optimizations of our simulator and adding some new functions to it, e.g. (1) quantum circuit testing and verification; (2) simulation of real circuits with quantum noise, and (3) simulation, debugging and verification of more sophisticated quantum algorithms and quantum programs (with control flows) \cite{yin16}. Another line of research is to develop more efficient simulation method using new mathematical and/or algorithmic tools like tensor network \cite{tensor network} or QuIDD \cite{quanSim}. We hope that our simulator can be extended to serve as a useful tool in the design, testing and validation of future quantum computer hardware and software. 

\section*{Acknowledgement}
We are very grateful to Gan Lin, Yu Haining, Zhang Wei, Shi Shupeng, Meng Hongsong, Yu Hongkun, Zhao Wenlai and the whole team at the National Super Computing Center in Wuxi for their kind helps. Special thanks go to Liu zhao, who has given us a lot of useful suggestions and assistance. This work is partially supported by the National Natural Science Foundation of China and the National Supercomputing Center in Wuxi.

\end{document}